\documentclass[aps,prb,twocolumn,showpacs]{revtex4}

\usepackage{boxedminipage}  
\usepackage{lscape}     
\usepackage{float}      
\usepackage[ansinew]{inputenc}  
\usepackage{epsfig}
\usepackage{subfigure}
\usepackage{graphicx}
\usepackage{amsmath}
\usepackage{amssymb}

\renewcommand{\d}[2]{\frac{#1}{#2}}
\newcommand{\pd}{\partial}

\begin{document}

\hyphenation{Brillouin}

\title{Hall conductivity in the presence of spin-orbit interaction and disorder}

\author{Louis-Fran\c cois Arsenault}
\altaffiliation{Present address: Département de Physique and RQMP,
Université de Sherbrooke, Sherbrooke, QC, Canada}
\affiliation{Department of Physics and Astronomy, Rutgers
University, Piscataway, NJ 08854-8019}
\author{B. Movaghar}
\altaffiliation{Present address: Department of Electrical and Computer Engineering,
Northwestern University, Evanston, IL, USA}
\affiliation{D\'epartement de G\'enie Physique and
Regroupement Qu\'eb\'ecois sur les Mat\'eriaux de Pointe (RQMP)\\
\'Ecole Polytechnique de Montr\'eal, C.P. 6079, Succursale
``Centre-Ville'', Montr\'eal (Qu\'ebec), H3C 3A7, Canada}

\date{\today}

\begin{abstract}
Starting from the Kubo formula, we expand the Hall conductivity
using a cumulant approach which converges quickly at high
temperatures ($k_BT >$ energy differences of initial and final
scattering states) and can be extended to low temperatures. The
theory can deal with the sign, the ordinary and the anomalous
contributions to the Hall effect. When applied to
include the spin-orbit interaction to first order, we recover what
is essentially the Karplus-Luttinger result for the anomalous
Hall effect. Contact is made to the Chazalviel and Nozières-Lewiner
formulae. A side-jump type formula is obtained by using an exact
application of linear response. We show that there exists an exact rigid
Hall current which is not a Fermi level property. We introduce a
relationship between mass and diffusivity which allows us to
generalize the theory to strong disorder and even introduce a
mobility edge. The formalism provides a systematic and practical way
of analyzing both ordinary and anomalous contributions to the
Hall conduction including the changes of sign, and in the presence
of serious disorder. As a byproduct of the method, we show that the anomalous Hall coefficient can vary with resistance to the power $n$, with
$1\leq n \leq 2$ depending on the degree of coherence.
\end{abstract}


\maketitle

\section{Introduction\label{sec:intro}}
The Hall conductivity of materials exhibits a wide and rich variety
of behavior. The interpretation is, in general, still very
difficult, even though, in principle, the information is contained
in the Kubo formula. This is true at least in linear response to the
applied electric field. The Kubo formula is therefore the starting
point of our analysis of Hall conductivity. We include the
spin-orbit interaction and discuss the so called anomalous
Hall effect (AHE). Previously, most work was focused on understanding
the sign change of the Hall coefficient in ordered and disordered
systems\cite{Movaghar_Cochrane1}, localization, and the Quantum Hall
effect (QHE)\cite{Ando}. More recently, the problem has been to
understand the effect of magnetism and of many body corrections on
the Hall effect. Considerable progress has been
made recently by integrating the spin-orbit coupling into the Bloch
wavefunction formalism of Kane in crystals, and applying these
wavefunctions and other first principle numerical methods\cite{Yao}
to study the AHE effect in magnetic materials\cite{Abolfath}. In
these papers, the emphasis is on order, and disorder is only
represented by a uniform lifetime. Most magnetically doped
semiconductors and magnetic alloys can, however, not really be
considered to be in the weak scattering regime, and therefore, in
the present approach, we have inverted this priority.
We emphasize the absence of Bloch symmetry rather than its presence\cite{Jungwirth1,Jungwirth2,Jungwirth3,Sinova1,Sinova2,Karplus_Luttinger}.\\
\\
The aim is to develop practical formulae which can deal with
disorder, the sign, the side jump Hall effect problem, and the Skew scattering/intrinsic Hall
effect problem.
\\
\\
The AHE is now a well established observation in magnets, and a
number of explanations have been proposed\cite{Jungwirth3}, and will
be discussed here. All explanations are based, in one way or
another, on the spin-orbit mechanism. Thus, it is accepted that
spin-orbit coupling causes the anomalous Hall conductivity
contribution.\\
\\
Much progress has been made recently on understanding the origin of
the AHE and this has generated exciting new physics. Traditionally
it was thought that only the spin-orbit \emph{Skew scattering
mechanism}\cite{Ballentine,Engel,Movaghar_Cochrane1}, gives rise to
a magnetism $(M_z)$ dependent Hall coefficient. This process depends
on the conductivity relaxation time or relaxation time squared,
depending on whether the skew scattering is itself rate determining
for conductivity or not. Traditionally, skew scattering is derived
as an extrinsic effect, i.e. it is due to impurity scattering, and
not to the host spin-orbit interaction. The intrinsic spin-orbit
interaction produced by the host crystal potential had been invoked
as a source of AHE by Karplus and Luttinger\cite{Karplus_Luttinger}
but later rejected by Smit\cite{Smit} who claimed that the intrinsic
effect is negligible. Earlier, it had been shown by Mott and
Massey\cite{Mott_Massey} that electron scattering from Coulomb
potentials is asymmetric with respect to spin direction, with spin
up going more to one side, and those with spin down more to the
other. When the electron gas is magnetized, there is a net
transverse Hall current. When the asymmetric scattering is at
impurity sites, this process is called skew scattering as mentioned
above, and has been discussed by several
authors\cite{Ballentine,Engel}. The important point about skew
scattering is that the net spin-orbit coupling at the impurity site,
can be strongly enhanced by the host crystal. This is really what
makes this process so important. For a detailed account of the history and
progress in understanding the AHE see Ref.~\onlinecite{Sinova1}.
\section{The Kubo formula for the longitudinal and transverse conductivity}
The frequency ($\omega$) dependant conductivity in linear response
to an electric field is usually written
as\cite{Movaghar_Cochrane1,Movaghar_Cochrane2,Roth}
\begin{equation}\label{kubo}
\sigma_{\mu\nu} =
\d{i\hbar e^2}{\Omega}\lim_{\delta \rightarrow
0}\sum_{\alpha,\beta}\d{\langle\alpha|v_{\mu}|\beta\rangle\langle\beta|v_{\nu}|\alpha\rangle}{\varepsilon_{\alpha}
- \varepsilon_{\beta} + \hbar\omega +
i\delta}\d{f(\varepsilon_{\alpha}) -
f(\varepsilon_{\beta})}{\varepsilon_{\beta} - \varepsilon_{\alpha}}.
\end{equation}
This form is general for any exact set of eigenstates
$|\alpha\rangle$ and energies $\varepsilon_{\alpha}$, spin summation
is implied. The $f(\varepsilon)$ are the Fermi functions and
$\Omega$ is the volume. The velocity operators $v_{\mu}$ are given
by Heisenberg's equation of
motion.\\
\\
For the general case of a material which need not be periodic, the spin orbit term in the Hamiltonian is
\begin{equation}\label{H_so}
H_{so}
= \d{\hbar}{4m^2c^2}\left( \nabla V(\textbf{r})\times\textbf{p}
\right)\cdot\boldsymbol{\sigma},
\end{equation}
where
\begin{equation}\label{pot_V}
    V(\textbf{r}) = \sum_n\d{eZ_n}{4\pi\varepsilon\varepsilon_0|\textbf{r}-\textbf{R}_n|} -eFx.
\end{equation}
In Eqs.~\eqref{H_so} and \eqref{pot_V}, $m$ is the bare electron mass, $c$ is
the speed of light, $\textbf{p}$ is the momentum operator and $\boldsymbol{\sigma}$
is the Pauli spin operator which is a vector containing
the Pauli's matrices i.e. $[\sigma_x, \sigma_y, \sigma_z]$. In Eq.~\eqref{pot_V}, $F$ is
the external applied electric field, $e$ the electric charge,
$Z_n$ is the effective local charge, $\varepsilon\varepsilon_0$ is the permittivity, $\textbf{r}$
and $x$ are positions operators for the charges and $\textbf{R}_n$ is
the position of the fixed ions that make the lattice. The velocity operators can then be
written as
\begin{equation}\label{vx_so}
    v_x = v_x^0 + \d{\hbar}{4m^2c^2}\Big[ \nabla_zV(\textbf{r})\sigma_y - \nabla_yV(\textbf{r})\sigma_z
    \Big],
\end{equation}
\begin{equation}\label{vy_so}
    v_y = v_y^0 - \d{eB_zx}{m} + \d{\hbar}{4m^2c^2}\Big[ \nabla_xV(\textbf{r})\sigma_z - \nabla_zV(\textbf{r})\sigma_x
    \Big],
\end{equation}
where we have used the Landau gauge for the vector potential,
$\textbf{A} = (0,B_zx,0)$. The $v_x^0$ and $v_y^0$ are
$-\d{i\hbar}{m}\d{\pd}{\pd x}$ and $-\d{i\hbar}{m}\d{\pd}{\pd y}$
respectively. The choice of the minus sign in $-eB_zx/m$ implies
that $e = -|e|$ since the kinetic part of the Hamiltonian for a
charge $q$ in a field is $T = 1/2m(\textbf{p} -
q\textbf{A})^2$.\\
\\
The spin dependent terms in Eqs.~\eqref{vx_so} and \eqref{vy_so} can
be important in magnets. We shall consider them explicitly in
Section~\ref{sec:spin-orbit}.\\
\\
The Hall effect is given by the antisymmetric part of the transverse
conductivity\cite{Matsubara,Ballentine}. To write the antisymmetric
Hall conductivity, we note that for non-zero frequency it is the
anti-hermitian part\cite{Ballentine} $\sigma_{xy}^a =
\d{1}{2}(\sigma_{xy} - \sigma_{yx}^*)$. In the dc limit we have
\begin{equation}\label{trans_conduc3}
    \sigma_{xy}^a = \d{\hbar e^2}{\Omega}\lim_{\delta\rightarrow 0}\sum_{\alpha,\beta}  -i\langle\alpha |v_x|\beta\rangle\langle\beta
    |v_y|\alpha\rangle\d{f(\varepsilon_{\alpha})-f(\varepsilon_{\beta})}{ \left(\varepsilon_{\beta} -
    \varepsilon_{\alpha}\right)^2 + \delta^2}.
\end{equation}
We shall now derive the Hall conductivity in a general way which
will include both the skew scattering and the intrinsic
contributions.
\section{The Hall effect using a Cumulant expansion of the Kubo formula}
One way to derive the contributions
to the Hall effect, with disorder present, is to rewrite the Kubo
formula using the Heisenberg equation of motion:
\begin{equation}\label{matrix_x}
    \langle\alpha |x|\beta\rangle = -i\hbar\d{\langle\alpha |v_x|\beta\rangle}{\varepsilon_{\alpha} -
    \varepsilon_{\beta}},
\end{equation}
which is true in any finite box (length $L$) without dissipation.Then Eq.~\eqref{trans_conduc3} becomes
\begin{equation}\label{trans_conduc4}
\begin{split}
    &\sigma_{xy}^a =\\ &\d{e^2}{\Omega}\lim_{\delta\rightarrow 0}\sum_{\alpha,\beta}\d{\Big(f(\varepsilon_{\alpha})-f(\varepsilon_{\beta})\Big)\left( \varepsilon_{\alpha} - \varepsilon_{\beta} \right)}{ \left( \varepsilon_{\beta} - \varepsilon_{\alpha} \right)^2 + \delta^2}\langle\alpha |x|\beta\rangle\langle\beta
    |v_y|\alpha\rangle.
\end{split}
\end{equation}
We drop the $a$ superscript for the Hall conductivity, but it will be implicit that we refer to the antisymmetric part unless otherwise mentioned. In order to demonstrate the cumulant technique, we
first consider the limits of high temperatures when $\left(
\varepsilon_{\beta} - \varepsilon_{\alpha} \right) < k_BT$ and, of
small magnetic fields $k_BT \gg \hbar\omega_c$. This is the limit
when the matrix elements are dominated by intraband scattering with
weak $B$-field. We expand the function
\begin{equation}\label{cumulant}
\begin{split}
    &f(\varepsilon_{\alpha})-f(\varepsilon_{\beta}) \approx \text{e}^{-\d{\varepsilon_{\alpha} - \varepsilon_f}{k_BT}} -
    \text{e}^{-\d{\varepsilon_{\beta}- \varepsilon_f}{k_BT}}\\ &= \left( \d{\varepsilon_{\beta} -
\varepsilon_{\alpha}}{k_BT}
\right)\text{e}^{-\d{\varepsilon_{\alpha}- \varepsilon_f}{k_BT}}-\d{1}{2!}\left(
\d{\varepsilon_{\beta} - \varepsilon_{\alpha}}{k_BT}
\right)^2\text{e}^{-\d{\varepsilon_{\alpha}- \varepsilon_f}{k_BT}} + \ldots\\
\end{split}
\end{equation}
and substitute Eq.~\eqref{cumulant} and Eq.~\eqref{matrix_x} into Eq.~\eqref{trans_conduc3}.
This gives us a cumulant expansion of the Hall conductivity in
powers of $\d{1}{k_BT}$ where one can use the operator identity
\begin{equation}\label{op_identity}
    \left( \varepsilon_{\beta} -
\varepsilon_{\alpha} \right)A_{\alpha\beta} =
\langle\alpha|[A,H]|\beta\rangle,
\end{equation}
where $[A,H]$ is the commutator of $A$ with the Hamiltonian $H$ and
$A_{\alpha\beta} = \langle\alpha|A|\beta\rangle$, to reduce and
evaluate the terms generated by the expansion.
\\
\\
The high $T$ expansion converges rapidly as soon as $k_BT$ is larger than the typical energy difference in the matrix elements. In particular, this is obviously true as soon as $k_BT$ exceeds the bandwidth. But, we can also have a more general expansion that may also be valid at low temperature though the justification in this case is more complicated. To obtain the more general cumulant expansion at any temperature, we observe that in Eq.~\eqref{trans_conduc3} the dominant region is around $\varepsilon_{\beta} \approx \varepsilon_{\alpha}$. Thus, we carry out a complete Taylor expansion of $f(\varepsilon_{\beta})$ around $\varepsilon_{\alpha}$ and obtain
\begin{equation}\label{low_T_cumulant}
    f(\varepsilon_{\alpha}) - f(\varepsilon_{\beta}) = \left(-\d{\pd f(\varepsilon_{\alpha})}{\pd\varepsilon_{\alpha}}\right)(\varepsilon_{\beta}-\varepsilon_{\alpha})+\ldots
\end{equation}
When we substitute Eq.~\eqref{low_T_cumulant} (stricly equivalent to Eq.~\eqref{cumulant} at high $T$) into the Kubo formula (Eq.~\eqref{trans_conduc4}), we obtain the first order term
\begin{eqnarray}
    \sigma_{xy}^{\{1\}} = -\d{e^2}{\Omega}\lim_{\delta\rightarrow 0}\sum_{\alpha,\beta}& &\left(-\d{\pd f(\varepsilon_{\alpha})}{\pd\varepsilon_{\alpha}}\right)\d{(\varepsilon_{\beta}-\varepsilon_{\alpha})^2 + \delta^2 - \delta^2}{(\varepsilon_{\beta}-\varepsilon_{\alpha})^2 + \delta^2}\nonumber\\ & &\times\langle\alpha |x|\beta\rangle\langle\beta |v_y|\alpha\rangle\nonumber
\end{eqnarray}
\begin{equation}
\begin{split}
 = &-\d{e^2}{\Omega}\Bigg[\sum_{\alpha}\left(-\d{\pd f(\varepsilon_{\alpha})}{\pd\varepsilon_{\alpha}}\right)\langle\alpha |xv_y|\alpha\rangle\nonumber\\ &-\lim_{\delta\rightarrow 0}\sum_{\alpha,\beta}\left(-\d{\pd f(\varepsilon_{\alpha})}{\pd\varepsilon_{\alpha}}\right)\d{\delta^2}{(\varepsilon_{\beta}-\varepsilon_{\alpha})^2 + \delta^2}\langle\alpha |x|\beta\rangle\langle\beta |v_y|\alpha\rangle \Bigg]\nonumber
\end{split}
\end{equation}
\begin{equation}\label{Hall_cumulant1}
    = -\d{e^2}{\Omega}\sum_{\alpha}\left(-\d{\pd f(\varepsilon_{\alpha})}{\pd \varepsilon_{\alpha}}\right)\langle\alpha|xv_y|\alpha\rangle,
\end{equation}
where we have used the identity $\lim_{\delta\rightarrow 0}\d{\delta^2}{(\varepsilon_{\beta}-\varepsilon_{\alpha})^2 + \delta^2} = \pi\delta (\varepsilon_{\alpha}-\varepsilon_{\beta})\lim_{\delta\rightarrow 0}\delta = 0$.\\
\\
Apart from this first order term (Eq.~\eqref{Hall_cumulant1}), we obtain an infinite series of higher order cumulant which can be generated via Eq.~\eqref{op_identity}. These higher order cumulants represent correction to the first order result. The first order term will turn out to have very simple physical interpretation. The corrections generated by the higher order cumulants resulting from Eq.~\eqref{low_T_cumulant}, and when evaluated at low temperatures, must be studied in detail as it is not immediately self-evident that they represent lower order corrections. This is done later in the  paper, but at this stage, it is already possible to note that the corrections represent, via Heisenberg's equation of motion, and from Eq.~\eqref{op_identity}, higher and higher time derivatives of the velocity operator. When using the effective mass Hamiltonian, one can show that beyond the second order cumulant, the corrections which are linear in the cyclotron frequency or magnetisation all scale with the disorder potential and spin-orbit coupling, and therefore are lower order corrections. From the mathematical structure and to keep the terms linear in $B_z$, it is thus essential therefore only to keep the second order cumulant at low temperatures. This will now be shown step by step as we proceed with the analysis of the various contributions to the Hall current.\\
\\
Let us also remember that in Eq.~\eqref{Hall_cumulant1}, $|\alpha\rangle$ is a magnetic field dependent exact eigenstate.\\
\\
In order to rewrite the matrix element in a simple way, we show in Appendix~\ref{appen_deriv_of_H}, that we can write the derivative of the eigenvalue with respect to the magnetic field as
\begin{equation}\label{deriv_H}
    \d{\pd\varepsilon_{\alpha}(B_z)}{\pd B_z}
    = -e\left\langle\alpha\left|xv_y\right|\alpha\right\rangle -
\d{g\mu_B}{2}\left\langle\alpha\left|\sum_i\sigma_z^i\right|\alpha\right\rangle,
\end{equation}
where $v_y$ does not contains the spin-orbit term
(Eq.~\eqref{vy_so}) contrary to $v_y$ in Eq.~\eqref{Hall_cumulant1}.
The spin-orbit term of $v_y$ will be treated later, in
Section~\ref{sec:v_y:SO}.\\
\\
The Hall conductivity only comes from the first term on the RHS of
Eq.~\eqref{deriv_H}, the orbital term. The way to handle this is to
introduce initially two different magnetic fields, one acting on the
orbital part $B_{orb}$ and one giving the Zeeman energies $B_z$. We
can thus rewrite Eq.~\eqref{Hall_cumulant1} as
\begin{equation}\label{Hall_cumulant_3}
    \sigma_{xy}^{\{1\}} = \d{e^2}{\Omega}\d{1}{e}\sum_{\alpha}\left(-\d{\pd f(\varepsilon_{\alpha})}{\pd\varepsilon_{\alpha}}\right)
\d{\pd\varepsilon_{\alpha}}{\pd B_{orb}},
\end{equation}
Only the derivative of the energy eigenvalues with respect to the orbital
field $B_{orb}$ gives the Hall conductivity. An advantage of this
representation is that the spin-orbit energy can now be treated in
first order perturbation theory, as we shall see later.\\
\\
The way disorder should be treated for numerical calculations is as follows. The Hall conductivity as given by the general
conductivity formula (Eq.~\eqref{kubo}) or the formulae obtained using the cumulant approach (Eq.~\eqref{Hall_cumulant_3} and equations thereafter) is evaluated for a specific configuration of disorder. The wavefunctions $|\alpha\rangle$ and energies $\varepsilon_{\alpha}$   are therefore the exact corresponding eigenstates and eigenvalues. Now the procedure is repeated for all the possible configurations and averaged with the appropriate weighting factors.\\
\\
Before examining the higher order terms (Appendices~\ref{second_order_appen} and \ref{third_order_appen}),
let us understand the significance of this result and compare with other well known approaches.
\section{Comparison to other theories}
\subsection{The Streda result}
We can rewrite Eq.~\eqref{Hall_cumulant_3} in the form
\begin{equation}\label{Hall_cumulant_lowT_1}
    \sigma_{xy}^{\{1\}} = -e\left[\d{\pd}{\pd
    B_z}\int_{-\infty}^E\rho(E',B_z)dE'  \right]_{E =
    \varepsilon_f},
\end{equation}
where $\rho(E',B_z)$ is the density of states with magnetic field.\\
\\
This expression is the quantum term of the Streda
formula\cite{Streda}. This author reduced the Kubo
formula to two terms called $\sigma_{xy}^I$ and $\sigma_{xy}^{II}$
(Eq.~\eqref{Hall_cumulant_lowT_1} above). The antisymmetric part of
the other term, $\sigma_{xy}^{Ia}$, should be contained in the
remainder of the cumulant expansion. Our term
(Eq.~\eqref{Hall_cumulant_lowT_1}) differs from
Streda's\cite{Streda} by a minus sign. The sign problem can be traced
in Ref.~\onlinecite{Streda} to one transformation (Eq.(11) of Ref.~\onlinecite{Streda}) where there should be a minus sign on the RHS.
\subsection{The classical limit}\label{sec:class:limit}
If, instead of Eq.~\eqref{matrix_x}, we write as in Ref.~\onlinecite{Datta}
\begin{equation}\label{matrix_x_classical}
    \langle\alpha |x|\beta\rangle = -i\hbar\d{\langle\alpha
|v_x|\beta\rangle}{\varepsilon_{\alpha} - \varepsilon_{\beta} +
\d{i\hbar}{\tau}},
\end{equation}
where $\tau$ is a lifetime, we can reduce the first term of the
cumulant expansion to the classical result. By adding a lifetime, we effectively assume that the electrons are subject to
resistive scattering processes. Using the effective mass Hamiltonian for the periodic part of the Hamiltonian
also allows us to replace $m$ with the effective $m^*$. The finite lifetime then represents all the scattering processes
that break the translational invariance of a Bloch electron with effective mass $m^*$. This includes disorder and electron-phonon scattering treated in the Born approximation. When working in this approximation, there is no configuration average to be done anymore.\\
\\
We consider only the term linear in $B_z$ in
Eq.~\eqref{trans_conduc4}, the one involving the $eB_zx/m$ part of
$v_y$ (see Eq.~\eqref{vy_so}), which gives the diagonal mass tensor
term. Then, Eq.~\eqref{trans_conduc4} together with Eq.~\eqref{matrix_x_classical} yields the cumulant,
\begin{equation}\label{cumulant_hall_classical_2}
\begin{split}
    \sigma_{xy}^{\{1\}} &= \d{e^2}{\Omega
    m}\Big(eB_z\Big)\sum_{\alpha}\left(-\d{\pd
    f(\varepsilon_{\alpha})}{\pd\varepsilon_{\alpha}}\right)\langle\alpha|x^2|\alpha\rangle\\
    &\approx \d{e^2}{\Omega m^*}\Big(eB_z\Big)\sum_{\alpha}\left(-\d{\pd
    f(\varepsilon_{\alpha})}{\pd\varepsilon_{\alpha}}\right)\sum_{\beta}\tau^2\left| \langle\alpha |v_x|\beta\rangle
    \right|^2.
\end{split}
\end{equation}
This leads to the very well known classical result
\begin{equation}\label{class_result_Fermi}
    \sigma_{xy}^{\{1\}} = \d{Ne^2\tau}{m^*}\d{eB_z\tau}{m^*} =
    \sigma_{xx}\d{eB_z\tau}{m^*},
\end{equation}
where $N = \rho(\varepsilon_f)m^*v_f^2$.
\section{The influence of the spin orbit coupling}\label{sec:spin-orbit}
Spin-orbit coupling introduces a number of new contributions to the
Hall current. New terms arise, due to the spin dependent velocities
from Eqs.~\eqref{vx_so} and~\eqref{vy_so}, and from the effect of
the spin-orbit interaction on the energy levels. We examine this
last effect first. Also, as pointed out in Section~\ref{sec:class:limit}, for the class
of problems where we have a periodic system + impurities, wherever the bare mass appears in the
following text(except when it comes from the spin-orbit
Hamiltonian), we can replace bare mass with the effective
mass ($m^*$) and drop the periodic part of the Hamiltonian
in the remaining analysis.
\subsection{The effect of the spin orbit interaction on the  Hall
current from the changes in the energy levels and wavefunctions}
Consider the first order cumulant result with the zero order
velocity operators (including the $-\d{eB_zx}{m}$ term). The
advantage of the cumulant expansion is that it allows us to analyze
a very complex phenomenon, the effect of the spin-orbit coupling on
the Hall conductivity, via the eigenstates. We write, to first order
in perturbation theory,
\begin{equation}\label{first_order_so_energy}
    \varepsilon_{\alpha}(B_z) = \varepsilon_{\alpha}^0(B_z)+
    \langle\alpha|H_{so}|\alpha\rangle.
\end{equation}
The spin orbit Hamiltonian being dependent upon the Pauli's matrices, we should remember that the state $|\alpha\rangle$ must now be a spinor (two components vector). The action of taking the bracket will leave a scalar and, as we use the same state at this order, only the $z$ component
survive.
\begin{equation}\label{deriv_first_order_so_energy}
\begin{split}
    \d{\pd\varepsilon_{\alpha}(B_z)}{\pd B_{orb}} &=
    -e\langle\alpha|xv_y|\alpha\rangle\\ &= \d{\pd\varepsilon_{\alpha}^0(B_z)}{\pd
    B_{orb}} + \sigma_z^{\alpha}\d{\pd}{\pd B_{orb}}\left\langle\alpha
    \left|\sum_i\lambda_il_{i,z}\right|\alpha\right\rangle,
\end{split}
\end{equation}
where
\begin{equation}\label{lambda}
    \lambda_n = \d{\hbar e Z_n}{4m^2c^2(4\pi\varepsilon\varepsilon_0)\left| \textbf{r} - \textbf{R}_n
    \right|^3},
\end{equation}
$\sigma_z^{\alpha} \equiv \langle\alpha |\sigma_z|\alpha\rangle$ and $l_{i,z}$ is the $z$ component of the orbital angular momentum at site $i$.\\
\\
The first term in Eq.~\eqref{deriv_first_order_so_energy} is the one we examined above
in the classical limit and is intuitively very attractive. The
magnitude of the Hall current per eigenstate is related to the
sensitivity of the energy level to an external magnetic field. Its
sign depends on whether the magnetic field increases or decreases
the energy of the eigenstate. In particular, it also follows that
the contribution of a localized state is negligible. In reality,
localized states should actually give exactly zero. It trivially follows that the zero is recovered  only after summing the remaining contributions in the cumulant series. Keeping only the first cumulant does not give the
exact result when the level in question is a localized level, with discrete energy levels. But  the first order  result is close to zero, and
therefore can be said to represent a good approximation. See
Appendix~\ref{second_order_appen} and \ref{third_order_appen} for
the analysis of the higher order
cumulants.\\
\\
For delocalized states, there is more information in the first term
of Eq.~\eqref{deriv_first_order_so_energy}. Normally, for weak
scattering, when $\varepsilon_f$ is near the top of the band, we
have the hole sign, because the magnetic field can only lower the
energy near the top of the band. Near the bottom of a band, we have
the electron sign because the magnetic field confines the carrier
and raises the energy of the electrons. This rule is also true for
disordered eigenstates. The effect of the magnetic field on the
energy levels can be evaluated in second order perturbation theory
in the presence of disorder .\\
\\
The anomalous term can now be studied by going to first order
perturbation theory in $B$-field with exact eigenstates. The term
$-\mu_B\textbf{L}\cdot\textbf{B}_{orb}$ generates
\begin{equation}\label{eigen_so}
    |\alpha\rangle = |\alpha_0\rangle +
    \sum_{\beta_0}\langle\beta_0|-\mu_B\textbf{L}\cdot\textbf{B}_{orb}|\alpha_0\rangle\d{|\beta_0\rangle}{\varepsilon_{\alpha_0} -
    \varepsilon_{\beta_0}}.
\end{equation}
Substituting Eq.~\eqref{eigen_so} in the second term of
Eq.~\eqref{deriv_first_order_so_energy}, keeping only the
$z$-components, we obtain terms in the energy which involve a factor
of the type
\begin{equation}\label{delta_g}
    \Delta g_{\alpha}^{zz} = \sum_{\beta}\d{\sum_i\lambda_il_{\alpha\beta}^{i,z}\sum_jl_{\beta\alpha}^{j,z}}{\varepsilon_{\alpha} -
    \varepsilon_{\beta}}.
\end{equation}
The sign of the anomalous process depends upon the sign of a
quantity which is closely related to the electron $g$-shift and
which itself can be electron-like or hole-like. Thus the first order
cumulant can be written as
\begin{equation}\label{first_Hall_so}
    \sigma_{xy}^{\{1\}} = \sigma_{xy}^{\{1\}(n)} + \d{e^2}{\Omega}\d{1}{|e|}\sum_{\alpha}\left(-\d{\pd f(\varepsilon_{\alpha})}{\pd\varepsilon_{\alpha}}\right)\mu_B\sigma_{z}^{\alpha}\Delta g_{\alpha}^{zz},
\end{equation}
Let us compare the relative magnitude of the two terms of
Eq.~\eqref{first_Hall_so}. At very low $T$ ( $-\d{\pd f}{\pd\varepsilon} = \delta(\varepsilon-\varepsilon_f)$ )
we can rewrite the Hall conductivity (Eq.~\eqref{first_Hall_so}) and the current is
\begin{equation}\label{y_current}
    J_y = \sigma_{xy}^{\{1\}(n)}F_x +
    e^2\rho(\varepsilon_f)\d{\hbar}{2m}\langle\sigma_z\rangle_{\varepsilon_f}\Delta
    g^{zz}(\varepsilon_f)F_x.
\end{equation}
With $\rho(\varepsilon_f) \sim 10^{45}$ /m$^3$J and $F_x = 10^4$ V/m
we have for the anomalous contribution $J_{y}^{an} = 10^3\Delta
g\langle\sigma_z\rangle F_x$ A/m$^2$, which is the same order of
magnitude as the normal process with $N = 10^{26}$/m$^3$, giving the
normal Hall current $J_{y}^n=10^6$ A/m$^2$ or
$10^5(\hbar\omega_c\tau)F_x$ A/m$^2$ where $\hbar\omega_c\tau \sim
10^{-3}$. In principle, the normal and anomalous terms can have
opposite signs. The results of Eq.~\eqref{first_Hall_so} and
Eq.~\eqref{y_current} are very elegant, and by writing $\Delta g =
2-g^*$ we have essentially recovered the Fermi level version of the
Chazalviel\cite{Chazalviel} and Nozières and
Lewiner\cite{Nozieres_Lewiner} result. Chazalviel\cite{Chazalviel}
computes the single carrier wave packet motion in an electric field
without using the Kubo linear response formalism. We also note that
in this form, the anomalous term, apparently has no
dependence on the relaxation time. Finally, and most importantly, it
can also be related  to the Karplus and
Luttinger\cite{Karplus_Luttinger} result in the limit of high
temperature when the Karplus and Luttinger energy gap between the
Kane Luttinger subbands is taken as $\Delta \sim k_BT$ and $\Delta g \sim
\d{\lambda_{so}}{\Delta}$.\\
\\
In the Karplus-Lutttinger\cite{Karplus_Luttinger} Bloch wavefunction
formalism, the anomalous Hall effect, even though it is
intrinsic, is a Fermi level property at low temperatures. In
Ref.~\onlinecite{Karplus_Luttinger}, the spin-orbit interaction is
treated in first order perturbation theory using Bloch functions. We
have also used first order perturbation, but the Fermi level
property, here, is a result of keeping the first order cumulant. An
interesting point is that Karplus and Luttinger did not use the Kubo
formula. They arrived at a similar expression, except that the
matrix elements are always interband matrix elements. The reason is
that their starting point is the Bloch function, so that in the
absence of an explicit treatment of disorder scattering, only
interband matrix elements are left when the matrix elements of
position and momentum operators are considered. Note that one way of
calculating the $g$-shift, when we have weak disorder, is to use the
Kohn-Luttinger wavefunctions. One can compute the $g$-shift $\Delta
g_{\textbf{k},\alpha}$ in the exact band
states\cite{Roth1959}.\\
\\
In summary, in this section, we have derived a result which can be related to the Karplus-Luttinger
intrinsic AHE\cite{Karplus_Luttinger} and we have made contact with the
Chazalviel\cite{Chazalviel}, Nozières and
Lewiner\cite{Nozieres_Lewiner} and Sinova\cite{Sinova1,Sinova2}
results using a simple and unified formalism.
\subsection{The Side Jump Hall current}\label{side_jump}
The new velocity terms generated by the spin-orbit coupling in
Eqs.~\eqref{vx_so} and~\eqref{vy_so} have been shown by Wölfle and
Muttalib\cite{Wolfle_Muttalib} to give rise to a term of a form
called the "side jump Hall effect" in the linear response Kubo
formula, which is
\begin{equation}\label{WM_side_jump}
    J_y = -Ne^2\langle\sigma_z\rangle\d{\hbar}{4m^2c^2}F_x.
\end{equation}
However, this result is obtained only after a complex diagrammatic
sum of potential scattering events. A similar result was derived previously and more simply from the same velocity term by Lyo and Holstein\cite{Lyo_Holstein} using scattering theory and originally by Berger\cite{Berger}. We shall now rederive a contribution to the Hall current which has the same structure but is more closely related to the Rashba effect\cite{Rashba}. It is derived
using linear response with the external field induced spin-orbit term in the Hamiltonian, below. The linear response analysis will show us how the presence of the lattice goes to modify the Hall conductivity even when we have disorder. Treated in linear response, the Rashba coupling term will give a side jump like Hall current. We do not use the cumulant approach or effective mass approximation because
it is easy to treat this term exactly. \\
\\
Consider again that part of the spin-orbit coupling which is itself
dependent on the applied external field $F_x$ (see Eqs.~\eqref{pot_V}). This term is a contribution to the total energy which directly depends on the external field. As part of the Hamiltonian, this
term creates a departure from equilibrium, and must therefore be
treated on the same footing as the usual electric potential $eF_xx$.
We therefore start from first principles, with the density matrix.
When we take all such terms in the Hamiltonian as the perturbation
$H_{pert}$, the change in the density matrix is given by
\begin{equation}\label{density_matrix}
    \Delta D_{\alpha\beta} = \langle\alpha
    |H_{pert}|\beta\rangle\d{f(\varepsilon_{\alpha})-f(\varepsilon_{\beta})}{\varepsilon_{\beta}-\varepsilon_{\alpha}},
\end{equation}
where
\begin{equation}\label{Hpert}
    H_{pert} = -eF_xx + \d{\hbar}{4m^2c^2}\left( \nabla V_{ext}(\textbf{r})\times\textbf{p}
    \right)\cdot\boldsymbol{\sigma},
\end{equation}
with $V_{ext}(\textbf{r}) = -eF_xx$.\\
\\
The second term of Eq.~\eqref{Hpert} involves the external applied
electric field, and one can use linear response and ask: what Hall
current does it produce in the presence of disorder? We can evaluate
the thermally averaged velocities in the usual way. We consider the
external-field-independent eigenstates, include disorder and the
Zeeman splitting. These states can therefore be picked to have
either spin up or spin down eigenstates in a chosen direction. Thus
the $y$-current produced by the second term from
Eq.~\eqref{density_matrix} and~Eq.~\eqref{Hpert} is given by (spin
projection in $z$-direction and keeping only the $y$-momentum term)
\begin{equation}\label{y_current_2}
\begin{split} &J_y =\\
&-\d{e^2}{\Omega}F_x\sum_{\alpha,\beta}\langle\beta
|v_y|\alpha\rangle\left\langle\alpha\left|\d{\hbar}{4m^2c^2}p_y\right|\beta\right\rangle\d{f(\varepsilon_{\alpha})-f(\varepsilon_{\beta})}{\varepsilon_{\beta}-\varepsilon_{\alpha}}\sigma_{z}^{\alpha}.
\end{split} \end{equation}
If we now use $mv_y = p_y$ and the sum rule (to be discussed in
detail in Section~\ref{subsec:sum_rule})
\begin{equation}\label{sum_rule}
    \d{1}{2m_{\alpha}}=\sum_{\beta}\d{|\langle\alpha
    |v_{\mu}|\beta\rangle|^2}{\varepsilon_{\beta}-\varepsilon_{\alpha}},
\end{equation}
where $\mu = x$ or $y$, applied to the $y$-operator, we have the
very simple and elegant result
\begin{equation}\label{Jy_new}
    J_y = -\d{e^2}{\Omega}\left( \d{\hbar}{4m^2c^2}
    \right)F_x\sum_{\alpha}f(\varepsilon_{\alpha})\d{m}{m_{\alpha}}\sigma_{z}^{\alpha},
\end{equation}
which apparently only depends upon the effective mass. This is true
as long as Eq.~\eqref{sum_rule} can be used to define effective
mass, i.e. if all incoherence is neglected. The interpretation of
this sum rule in the Kubo formula context is not trivial. If we
include the entire infinite spectrum in the sum of
Eq.~\eqref{sum_rule}, then we have the trivial result $m_{\alpha} =
m$, and we  obtain a contribution which looks exactly like the Berger\cite{Berger} and Lyo and Holstein\cite{Lyo_Holstein} side jump Hall current but is clearly based on a different logic (Rashba term). Note that in our theory, the spin polarization is now
summed over the entire band and is not just the fermi level spin.\\
\\
In the framework of a finite band model, one can interpret
Eq.~\eqref{sum_rule} as the \emph{effective mass}. The spin-orbit
terms associated with the lattice can be included in
Eq.~\eqref{sum_rule}. In the weak disorder limit, one could compute
Eq.~\eqref{sum_rule} using the Kane method. One can see that, in the
Kane model, Eq.~\eqref{sum_rule} is indeed the effective mass. The
effective mass correction in Eq.~\eqref{Jy_new} can, then in
principle, increase the current up to two orders of magnitude (InSb
for example). But the side jump (Eq.~\eqref{Jy_new}) is, even with
effective mass, for extended states, much smaller than the
Karplus-Luttinger contribution. We have $J_y =
10^{-29}\langle\sigma_z\rangle NF_x$ A/m in 2-dimensions. In
3-dimensions, with the same numbers we have
$10^{-3}F_x\langle\sigma_z\rangle(m/m^*)$ A/m$^2$ compared to
$10^3F_x\langle\sigma_z\rangle\Delta g$ A/m$^2$ from
Eq.~\eqref{y_current}. Note that the huge Bloch function enhancement
evaluated in various forms by Chazalviel\cite{Chazalviel},
Fivaz\cite{Fivaz} and Berger\cite{Berger} and which make the side jump term
important does not appear in Eq.~\eqref{Jy_new}. Chazalviel, for
example, used the Heisenberg commutator for $v_y$ and Kane
wavefunctions to derive the Hall velocity and then derives a similar expression for the Hall current using Drude theory.
We have derived an expression which is similar in structure to what is called the side jump Hall current in the literature. The derivation we have used is however not the same as that of Berger\cite{Berger}, Lyo and Holstein\cite{Lyo_Holstein}and Wölfle and Mutallib\cite{Wolfle_Muttalib}. In our formula, the spin-orbit coupling can be enhanced by the lattice, but the effect is relatively small, and directly related to the effective mass lowering (see Eq.\eqref{Jy_new}). The so-called side jump theories\cite{Berger,Lyo_Holstein} in which the mechanism is due to the spin-orbit scattering induced sideways jump at impurity potentials, and the associated large enhancements caused by Bloch functions, have not been recovered using the present Kubo formula method.\\
\\
To complete the analysis of Eq.~\eqref{Jy_new} we need a discussion
of the sum rule of Eq.~\eqref{sum_rule} and we will defer this to
Section~\ref{subsec:sum_rule} because a similar term is encountered
in the next section.
\subsection{The effect of the spin dependent velocity on the Hall current: The terms which are due to the internal
potentials}\label{sec:v_y:SO} Let us consider the contributions to
the Hall current which results from including the contribution of
the remaining spin-orbit velocity terms (Eq.~\eqref{vx_so}
and~\eqref{vy_so}) in the Kubo formula. These now involve the
lattice potentials as sources of \emph{velocity}. For Coulomb
potentials, we have
\begin{equation}\label{mat_el}
\begin{split}
    &\langle\alpha |v_x|\beta\rangle\langle\beta |v_y|\alpha\rangle
    \rightarrow\\
    &\left\langle\alpha\left|\sum_n\d{eZ_n\hbar}{4m^2c^2(4\pi\varepsilon\varepsilon_0)}\d{y-Y_n}{|\textbf{r} -
    \textbf{R}_n|^3}\sigma_z\right|\beta\right\rangle\langle\beta
    |v_y|\alpha\rangle\\
    &\approx \sum_{\eta}\left\langle\alpha\left|\sum_n\varsigma\d{1}{|\textbf{r} -
    \textbf{R}_n|^3}\right|\eta\right\rangle\left\langle\eta\left|y\sigma_z\right|\beta\right\rangle\langle\beta
    |v_y|\alpha\rangle\\
    &\approx \left\langle\alpha\left|\sum_n\varsigma\d{1}{|\textbf{r} -
    \textbf{R}_n|^3}\right|\alpha\right\rangle\left\langle\alpha\left|y\sigma_z\right|\beta\right\rangle\langle\beta
    |v_y|\alpha\rangle,
\end{split}
\end{equation}
where $\varsigma \equiv
\d{eZ_n\hbar}{4m^2c^2(4\pi\varepsilon\varepsilon_0)}$.\\\\ Two
approximations were made here. $Y_n$ takes alternatively positive
and negative values and the term involving it would be zero if
$\textbf{r}$ were not present. But, even if we include $\textbf{r}$,
it will always give a smaller contribution compared to the first and
we therefore neglect it. Second, $\d{1}{|\textbf{r}
-\textbf{R}_n|^3}$ is local and therefore cannot couple different
sites but, in tight-binding for example, could couple different
orbitals at the same sites. We considered that the main contribution
comes from the matrix element taken between the same eigenstate and
this is why we considered only the $|\eta\rangle =
|\alpha\rangle$ term in the previous equation. One has to note that the
original integrals in Eq.~\eqref{mat_el} are convergent but when one breaks them up, then the integral $\langle\alpha|\d{1}{|\textbf{r}-\textbf{R}_n|^3}|\alpha\rangle$ is strictly speaking not convergent because one has taken
one position term $y-Y_n$ out of it. The underlying assumption which allows us to make this decoupling  is that, the orbit radius is never allowed to be smaller
than the effective atomic orbit of the valence state so that the cubic singularity does not occur.\\
\\
This term gives rise to a contribution in the first cumulant. To
obtain it, we start from Eq.~\eqref{trans_conduc3} instead of
Eq.~\eqref{trans_conduc4}. We substitute Eq.~\eqref{mat_el} in
Eq.~\eqref{trans_conduc3} and use Eq.~\eqref{matrix_x} to transform
$\langle\alpha|y|\beta\rangle$ to $\langle\alpha|v_y|\beta\rangle$.
The result is
\begin{equation}\label{new_contri_Hall}
   \sigma_{xy} = \d{\hbar^2e^2}{\Omega}\sum_{\alpha}\left(-\d{\pd f(\varepsilon_{\alpha})}{\pd
    \varepsilon_{\alpha}}\right)\Gamma_{\alpha}\d{\pd}{\pd\varepsilon_{\alpha}}\left(\d{1}{2m_{\alpha}}\right)\sigma_z^{\alpha},
\end{equation}
where
\begin{equation}
    \Gamma_{\alpha}=\left\langle\alpha\left|\sum_n\d{eZ_n\hbar}{4m^2c^2(4\pi\varepsilon\varepsilon_0)}\d{1}{|\textbf{r} -
    \textbf{R}_n|^3}\right|\alpha\right\rangle.
\end{equation}
Again we have used Eq.~\eqref{sum_rule}. Following
Datta\cite{Datta}, we will assume that, in a crystal, the sum rule
of Eq.~\eqref{sum_rule} is indeed the effective mass. The sum rule
(Eq.~\eqref{sum_rule}) will be discussed in detail in
Section~\ref{subsec:sum_rule}. Note that
if we use Eq.~\eqref{matrix_x_classical} with broadening, then an exact result can be obtained by taking the derivative with respect
to the broadening $\left(\d{i\hbar}{\tau}\right)$ instead of the energy in Eq.~\eqref{new_contri_Hall}. This relation will be used to derive Eq.~\eqref{def_deriv_eff_mass} for the strong scattering limit.
\subsection{The problem of the sum rule of Eq.~\eqref{sum_rule}}\label{subsec:sum_rule} There is a certain
arbitrariness in the use of the sum rule of Eq.~\eqref{sum_rule}
which we should clarify. We note that, if we use the standard
transformation of Eq.~\eqref{matrix_x}, then it follows, in principle, that when we
sum over the entire real spectrum of the Hamiltonian, we obtain the
free electron mass on the LHS. This is simply a consequence of the
trivial identity $p_xx-xp_x = -i\hbar$ with $p_x = mv_x$.\\
\\
However, if we follow Datta\cite{Datta}, and use the linear response
density matrix to compute the \emph{acceleration} of a particle in
an electric field,  then we obtain the effective mass, in the sense
of Newton's law, as given by Eq.~\eqref{sum_rule}. Consequently,
this gives the absurd result that the accelerating particle is
always free, irrespective of what its initial state is. In effect,
the trivial result signifies that if we wait long enough, then even
a strongly bound electron will eventually be free in an electric
field. That is, this result (mass is free mass), would represent the
very long time behavior, when the history of the particle is
irrelevant, and its acceleration in a constant field is truly
dominated by what happens when it has reached its final free state.
Datta\cite{Datta} concluded that he should use a finite band in the
evaluation of Eq.~\eqref{sum_rule}, and then the LHS is indeed the
effective mass in the sense of the tight-binding
band structure, for example.\\
\\
The solution of this problem, in general, seems to be that, in a
transport situation, where electrons are injected at one end and
absorbed at the other, the sum can only run over that part of the
spectrum which is accessible to the carrier in its lifetime, i.e.
for which $\hbar\sum_{\beta}|\langle\alpha
|v_x|\beta\rangle|^2\delta(\varepsilon_{\beta} -
\varepsilon_{\alpha} - \hbar\omega)$ is finite. In weak scattering,
the particle \emph{lives} in energy levels near the Fermi level
which have an effective mass, because it relaxes and emits energy to
the lattice. In a strong scattering situation, the kinetic energy of
the carrier can be of the same order as the scattering energy
uncertainty $\d{\hbar}{\tau}$. So here, we can relate the sum on the
RHS of Eq.~\eqref{sum_rule} directly to the quantum diffusivity
$D_{\alpha}^0$ (see Appendix~\ref{strong_scatt}). We proposes
therefore, in the strong scattering limit, where Bloch's theorem
does not apply, to define the effective mass, in the sum rule of
Eq.~\eqref{sum_rule} by the relation
\begin{equation}\label{eff_mass_strong_scatt}
    \d{1}{m_{\alpha}} = 2c_{\alpha}\d{D_{\alpha}^0}{\hbar},
\end{equation}
where $c_{\alpha}$ is a constant $\sim \left\langle
\d{(\varepsilon_{\alpha}-\varepsilon_{\beta})\tau_{\beta}}{\hbar}
\right\rangle_{\beta}$ which carries a sign and is averaged over the
band (See Appendix~\ref{strong_scatt} for a formal representation).
It is of order 1 when the energy $\varepsilon_{\alpha}$ is near the
bottom of the band or in a rapidly changing region of the DOS. We
shall henceforth absorb this constant in the definition of an
effective diffusivity, $D_{\alpha}$.
Equation~\eqref{eff_mass_strong_scatt} is exact
(Appendix~\ref{strong_scatt}).\\
\\
In the random phase limit, $c_{\alpha}$ is a relatively weak
function of energy and can be treated as a constant. The unit of
time is the scattering time. Acceleration with strong disorder is
therefore drift velocity divided by scattering time. With the same
definition, in the semi-classical limit we therefore have
\begin{equation}\label{def_deriv_eff_mass}
    \d{\pd}{\pd\varepsilon_{\alpha}}\left(\d{1}{2m_{\alpha}}\right)
    \sim -\d{D_{\alpha}\tau}{\hbar^2}.
\end{equation}
It follows that a localized initial state $|\alpha\rangle$ has no
acceleration without phonons, i.e. at zero temperature, its
effective mass in the sense of Eq.~\eqref{sum_rule} is infinite. If
one evaluates what diffusivity one needs to reproduce the electron
mass using Eq.~\eqref{eff_mass_strong_scatt}, one has $D \sim 1$
cm$^2$s$^{-1}$ which is not a small value in a disordered system. We
now have a way of interpreting terms involving the effective mass
$m_{\alpha}$. For example, for strong disorder, expressions of the
form $\sum_{\alpha}\d{1}{m_{\alpha}}\d{\pd^2f(\varepsilon_{\alpha})}{\pd\varepsilon_{\alpha}^2}$, as encountered in Appendix~\ref{second_order_appen} for the second
order cumulant can be written as
\begin{equation}\label{low_T_corres}
-\int\d{D(\varepsilon)\rho(\varepsilon)}{\hbar}\d{\pd}{\pd\varepsilon}\Big[\delta(\varepsilon-\varepsilon_f)\Big]
=
\d{1}{\hbar}\d{\pd\sigma(\varepsilon)}{\pd\varepsilon}\Bigg|_{\varepsilon_f},
\end{equation}
with $\sigma(\varepsilon) = \rho(\varepsilon)D(\varepsilon)$, where $\sigma(\varepsilon)$ is the energy dependent conductivity
with universal scaling properties near mobility edges\cite{Ando}.
\subsection{The resistance dependence of the anomalous term}
Experimentally one is normally interested in the resistance
dependence of the AHE and this is the criterion used to discriminate
between possible models. For example, the observed resistance
independence of the experimental anomalous Hall conductivity in
dilute magnetic semiconductors is taken as proof that the AHE is
intrinsic. However, it follows from the present analysis, that the
apparent scattering-time independence of the anomalous Hall
conductivity, as derived here in the form of
Eq.~\eqref{first_Hall_so}, and by Karplus and
Luttinger\cite{Karplus_Luttinger}, and by Jungwirth \emph{et
al}.\cite{Jungwirth3} constitutes a very special limit. The
scattering time independence of the $g$-shift (Eq.~\eqref{delta_g})
follows when the matrix elements are dominated by interband
processes where the energy differences are $\gg \d{\hbar}{\tau}$,
(see Ref.~\onlinecite{Karplus_Luttinger} for example). But this
implies that the basic band structure is Bloch-like, with
well-defined semiconductor bands. This is obviously not always
the case, and there can be many situations where disorder and band
crossings wash out the Kane subband gaps, and give rise to
arbitrarily small energy denominators in which the lifetime directly
enters the AHE as well. Thus, the $g$-shift (Eq.~\eqref{delta_g})
can very well involve the conductivity scattering time.\\
\\
The assumption of Bloch functions is made by Jungwirth \emph{et
al}.\cite{Jungwirth3}, and the resistance independence of their
result also rests on the existence of well defined subbands, and the
sums run over all occupied levels. Thus we conclude that for strong
spin-orbit scattering and weak disorder scattering, the anomalous
Hall conductivity will not depend strongly on the resistance, and
this then accounts for some of the experimental observations on the
AHE\cite{Ruzmetov}. In contrast, for weak spin-orbit coupling and
strong disorder, the anomalous Hall conductivity can vary as the
scattering time $\tau^n$, $0 < n < 2$. The Bloch matrix
elements\cite{Chazalviel} are not appropriate in the strong disorder
limit, and the present approach, though perturbational, is more
appropriate.
\section{Discussion}
The Kubo formula was expanded in a cumulant expansion which
converges quickly at high temperatures. One can also use the series expansion
at lower temperatures, provided one examines the higher order contributions for convergence. Working with the effective mass Hamiltonian, and keeping only the contributions   to first order in magnetic field allows one  to stop the expansion after the second order cumulant . The higher order cumulant  contributions are examined  in the Appendices~\ref{second_order_appen} and \ref{third_order_appen}.\\
\\
The second order cumulant derived in Appendix~\ref{second_order_appen}, also has a very interesting
structure. It gives terms which scale as the derivative of the
density of states at the Fermi level, both for the normal, and the
anomalous contributions. Previously, Bush and Guentherodt\cite{Bush_Guentherodt} had suggested that the sign of the Hall effect in disordered materials scales as the sign of the derivative of the density of states at the fermi level (elecron like for increasing and vice versa). In the present theory, we have actually successfully identified the quantities that determine the sign of the Hall effect. Indeed, we have made the very interesting observation that the sign of the normal term scales, as expressed in
the first term, related to the derivative of the energy with respect to
$B$-field. In the second cumulant contribution, the sign is related to the derivative of the density of
states at the Fermi level (at low $T$). Interestingly the two terms have opposite trends because an increasing density of states actually gives the hole sign, and not as one would intuitively expect and as Bush and Guentherodt\cite{Bush_Guentherodt} suggest, the electron sign. This is a truly remarkable result when applied to the strong scattering limit, because normally in weak scattering, the first term is dominant and that gives the intuitive result from Ref.~\onlinecite{Bush_Guentherodt}. But in very strong scattering, for example in the amorphous limit, and in the region near the mobility edge, it may happen that the second term dominates. If this happens, then we have a sign anomaly because the increasing density of  states at $\varepsilon_f$ for n-doping gives the hole sign and vice versa. This is exactly what is observed in the band edge of doped amorphous silicon\cite{Overhof_Thomas}. This observation merits a more detailed investigation which goes beyond the scope of this paper.\\
\\
In disordered systems, one can use the CPA (Coherent
Potential Approximation)\cite{Velicky} to describe the disordered
band structure for example\cite{Arsenault1,Arsenault2} and get
explicit results for the sign of the Hall effect. It turns out that
there is no simple rule for the sign of the Hall coefficient in CPA
either, but at the band edges, we do indeed have the same behavior
as predicted here. This has been discussed in detail in
Refs.~\onlinecite{Movaghar_Cochrane1} and \onlinecite{Arsenault2}.
In Ref.~\onlinecite{Movaghar_Cochrane1}, the case of an impurity
band is also considered.\\
\\
The disadvantage of the cumulant method is that the low temperature
limit has to be examined with care for convergence. In situations
with Bloch symmetry where the dominant matrix elements are on the
same energy shell, this is no problem. But in general, with
disorder, there is, in the  present formulation, in lowest order in
magnetic field $B_z$, unfortunately, an infinite number of terms.
This seems a big problem at first, but then it resolves itself. The
resolution of the difficulty is most obvious when we apply the first
order term in the limit that the states are localized at
$\varepsilon_f$. We obtain a Hall conductivity which is small, but
non-zero. This small contribution must be canceled by the remaining
linear terms in the series. Nevertheless  the
approximation is still good because it gives a negligible contribution to the Hall current,  knowing that the exact result should be strictly zero.
Eqs.~\eqref{sigma_xy_first_comm} and \eqref{sigma_xy_third_order}
have terms which scale as $B_z$. They are generally smaller than the
first order contributions we derived in
Section~\ref{sec:spin-orbit}. However, they involve higher
derivatives of the Fermi function and can be dangerous to handle at
low $T$. One may infer that, if the system has a density of states
and scattering times which are only weak functions of energy at the
Fermi level, these higher order cumulant terms are negligible. If
the density of states is a strong function of energy, the expansion
will not converge so easily. Indeed, near the band edges, there will
be mobility edges and localized levels which must give rise to a
null result without phonons, but the null result must be arrived at
by cancellation of many, albeit, small
contributions.\\
\\
At high temperatures, the higher order terms cause no problem and
can be neglected. At any temperature we may conclude
that a very good approximation is obtained by keeping only the first
and second cumulants, the spin orbit velocity contribution, and the external field induced spin orbit term. The final approximate formula for the Hall
conductivity thus becomes
\begin{widetext}
\begin{equation}\label{tot_conduc}
\begin{split}
    \sigma_{xy} = &\d{e^2}{\Omega}\sum_{\alpha}\left(-\d{\pd f(\varepsilon_{\alpha})}{\pd\varepsilon_{\alpha}}\right)\left[ \d{-1}{|e|} \d{\pd\varepsilon_{\alpha}}{\pd B_{orb}} + \d{\hbar}{2m|e|}\sigma_z^{\alpha}\Delta
    g_{\alpha}^{zz}
    \right] +
    \d{e^2}{2!\Omega}\sum_{\alpha}\d{\hbar}{2m_{\alpha}}\d{\pd^2f(\varepsilon_{\alpha})}{\pd\varepsilon_{\alpha}^2}\left[ \d{\hbar |e|B_z}{m} + \hbar\Gamma_{\alpha}\sigma_z^{\alpha} \right]\\
    &-
    \d{e^2}{\Omega}\d{\hbar}{4m^2c^2}\sum_{\alpha}f(\varepsilon_{\alpha})\d{m}{m_{\alpha}}\sigma_z^{\alpha}
    + \d{e^2\hbar^2}{\Omega}\sum_{\alpha}\left(-\d{\pd f(\varepsilon_{\alpha})}{\pd\varepsilon_{\alpha}}\right)\Gamma_{\alpha}\d{\pd}{\pd\varepsilon_{\alpha}}\left(\d{1}{2m_{\alpha}}\right)\sigma_z^{\alpha}.
\end{split}
\end{equation}
\end{widetext}
We may call Eq.~\eqref{tot_conduc} the weak-to-intermediate
scattering Hall conductivity. When we evaluate the cumulants,
one can use the effective mass Hamiltonian so that the periodic potential is not part of the Hamiltonian with which the higher order commutators Eq.\eqref{op_identity} are to be evaluated. This means that the only terms which contribute above the $2^{nd}$ order cumulant,  and which scale linearly with $B_z$, will be those which depend on the disorder and spin-orbit part of the Hamiltonian and this makes the
approximation of only keeping up to second order very accurate. Each of the four additive terms of Eq.~\eqref{tot_conduc} will
now be discussed and a simple interpretation given.\\
\\
The first expression in the bracket of the first term has been
discussed and is easy to interpret, but it is not completely trivial
to see that it simply reduces to the classical result
(Eq.~\eqref{class_result_Fermi}) if we used
Eq.~\eqref{matrix_x_classical} or Bloch functions in the Kubo
formula with a constant lifetime. In the pure quantum limit it gives
the Streda result but it is actually the \emph{normal} Hall effect.
The way this term should be handled depends on the problem in
question. In weak scattering it again gives
Eq.~\eqref{class_result_Fermi}, with a free electron mass. In the
tight-binding representation, one can evaluate it using second order
perturbation theory in the magnetic field dependent term in the
Hamiltonian.\\
\\
The second part of the first bracket involves the $g$-shift of the
delocalized levels above the mobility edge. In comparison, the
localized $g$-shift is negligibly small. In the weak scattering
limit, the $g$-shift can be evaluated using the Kane-Luttinger
wavefunctions\cite{Karplus_Luttinger}. This is given by Roth
\emph{et al}\cite{Roth1959}. In an incoherent situation, a disordered system with no Bloch symmetry and
strong scattering , we should use $m_{\alpha} =
\d{\hbar}{D_{\alpha}}$ as the effective mass.\\
\\
For the second term, the one involving a second derivative of the
Fermi function, one can, using the $m_{\alpha} =
\d{\hbar}{D_{\alpha}}$ approximation, obtain integral products of
the type $\sigma(\varepsilon) = \rho(\varepsilon)D(\varepsilon)$ as
shown in Eq.~\eqref{low_T_corres} which are energy dependent
conductivities, and which obey well known universal scaling
relations near the mobility edges.\\
\\
The third term was shown to be due to the effect of the
external-field-induced spin-orbit energy (Section~\ref{side_jump}).
With $m_{\alpha} = m^*$, this term reduces to the same form as the so called side jump contribution. Our theory shows that it can be enhanced via
a small effective mass and even extended to apply to strong
scattering via
Eq.~\eqref{eff_mass_strong_scatt}.\\
\\
Considering the second and third terms, we propose that, to a good
approximation, in most situations where Bloch functions cannot be
used, we may replace $m_{\alpha} = \d{\hbar}{D_{\alpha}}$, once the
sign has been determined via Eq.~\eqref{sum_rule}. This accounts for
localized states if any are present, because
$D_{\alpha}$ is zero.\\
\\
The last term in Eq.~\eqref{tot_conduc} is due to the spin-orbit
contribution to the velocity operators via the internal potentials
from Eqs.~\eqref{vx_so} and \eqref{vy_so}. This also vanishes for
localized states. Above the mobility edge, this contribution is
comparable to the second expression in the bracket of the second
term. This can be seen using the strong scattering correspondence (Eq.~\eqref{def_deriv_eff_mass}). The last term in
Eq.~\eqref{tot_conduc} then reduces to a term which resembles the
usual skew scattering term\cite{Engel} provided we interpret the skew scattering rate for Coulomb spherical potentials as
\begin{equation}\label{skew_time}
    \d{1}{\tau_s^{\alpha}} = \d{\hbar
    e}{4m^2c^2(4\pi\varepsilon\varepsilon_0)}\left\langle\alpha\left|\sum_n\d{Z_n}{|\textbf{r}- \textbf{R}_n|^3}\right|\alpha\right\rangle.
\end{equation}
Thus, for disordered systems, the last term of Eq.~\eqref{tot_conduc} term can
be written
\begin{equation}\label{skew_sigma}
    \sigma_{xy}^{skew} = -\d{\hbar e^2}{\Omega m}\sum_{\alpha}\left(-\d{\pd
    f(\varepsilon_{\alpha})}{\pd\varepsilon_{\alpha}}\right)\d{\tau}{\tau_s^{\alpha}}\d{mD_{\alpha}}{\hbar}\sigma_z^{\alpha}.
\end{equation}
Note that the sum runs over all the potentials in the lattice. So
the skew scattering and $g$-shift terms include both extrinsic and
intrinsic contributions. Note that very recently,
Chudnovsky\cite{Chudnovsky}, using a different approach, and for
spin Hall effect, also obtained a term where all the potentials are
included. In the form of Eq.~\eqref{skew_sigma}, appropriate for
disordered systems with no Bloch symmetry, the Bloch enhancement does not appear. If we neglect the host
spin-orbit coupling and only include the impurities, then both
$g$-shift and skew term are, by definition, extrinsic contributions,
and the mass in Eq.~\eqref{skew_time} is necessarily the effective
mass. It is the effective mass particle which generates the impurity
spin-orbit magnetic field.\\
\\
Let us now re-examine the question of the theoretical resistance
dependence of the AHE. From the above analysis we note that this all
has to do with the way we treat the matrix elements, and at what
stage we introduce incoherence and lifetime. This can already be
seen in the first term, which can be treated as a quantum effect, as
in Streda\cite{Streda} or in the semi-classical limit. The same is
true for the spin-orbit terms. Thus, if we keep to the notion of
effective mass, we have the quantum result. If we use the
transformation of Eq.~\eqref{eff_mass_strong_scatt}, which involves
the diffusivity, then we have the connection with conductivity. As
an example, we can take the first two, dominant, terms in
Eq.~\eqref{tot_conduc} and write them, using the definition of the $g$-shift given by Roth \emph{et al}.\cite{Roth1959} (we
assume $\Delta$ and $E_g$ to still be defined), in the incoherent limit as:
\begin{equation}\label{sigxy3}
\begin{split}
&\sigma_{xy}^{(1)} = \sigma_{xx}\left(\d{eB_z\tau}{m^*}\right)
 +\\ &\d{e^2}{\Omega}\sum_{\alpha}\left(-\d{\pd f(\varepsilon_{\alpha})}{\pd
 \varepsilon_{\alpha}}\right)\d{\hbar}{m}\sigma_z^{\alpha}\left[ -\left(m\d{D_{\alpha}}{\hbar} - 1\right)\d{\Delta}{3E_g + 2\Delta}
 \right],
\end{split}
\end{equation}
where we have used Eq.~\eqref{eff_mass_strong_scatt} in the $g$-shift term.\\
\\
Now, we see that what was a pure quantum term with effective mass,
has, in this limit, become a term which depends on the diffusivity.
The AHE Hall coefficient can, it seems, change from a linear scaling with resistance in Eq.\eqref{sigxy3} to a resistance squared
(relaxation time squared) behavior if
$\d{D_{\alpha}}{\hbar}$ becomes $m_{\alpha}^{-1}$, as is observed in
diluted magnetic semiconductors\cite{Ruzmetov}. The
experimentally observed $R_A \propto \rho_{xx}^2$ dependence
implies, therefore, that the effective mass concept remains valid in
these systems.\\
\\
We have focused our attention on the anomalous contributions, and
how they undergo a transformation, in going from the weak scattering
to the strong scattering limit. A similar change must occur for the
first term of Eq.~\eqref{sigxy3}. Here too, we must replace the
Drude $\sigma_{xx}$ by
\begin{equation}\label{replace1}
    \sigma_{xx} = e^2\int d\varepsilon\left(-\d{\pd
    f(\varepsilon)}{\pd\varepsilon}\right)\rho(\varepsilon)D(\varepsilon)
\end{equation}
and the band mobility term $\d{e\tau B_z}{m^*}$ by
\begin{equation}\label{replace2}
    \d{e\tau B_z}{m^*} = eB_z\d{D}{\langle\varepsilon\rangle},
\end{equation}
at low temperatures. The notation in terms of diffusivity $D$ is
valid in the weak scattering limit too, but now one can see what
happens as we approach the mobility edge using standard localization
theory and
$\sigma(\varepsilon) = \rho(\varepsilon)D(\varepsilon)$.\\
\\
We have shown that the AHE Hall coefficient $R_A$ can vary with
resistance $\rho_{xx}^n$ with $1\leq n \leq 2$ depending on the
degree of coherence. The essential point seems to be the way one
treats the sum rule given by Eq.~\eqref{sum_rule}.
This was already a problem for Datta\cite{Datta} in his analysis of
the Hall effect in his 1980 paper. Eq.~\eqref{sum_rule} may be
treated as the well known f-sum rule, but it is clearly not
realistic to sum over an infinite excited state spectrum, without
taking into account the finite lifetime of the states. The question
then becomes:  when is the sum on the RHS of Eq.~\eqref{sum_rule}
$\d{1}{2m^*}$ and when is it closer to $\d{D}{\hbar}$ which is the
sum evaluated in the semi-classical limit, and also derivable from
$\langle\alpha |x^2|\alpha\rangle \sim D_{\alpha}\tau$? The
interesting and important side of this last result, is that it gives
the correct null result for transport in localized states, and
therefore must be the correct approximation near the mobility
edges.\\
\\
We have derived a side-jump-like contribution to the Hall current using linear response to the (Rashba) spin-orbit interaction\cite{Rashba} which is produced by motion in the external potential. Our result is closely related to the Rashba current\cite{Rashba} and is in principle, exact, up to the evaluation of the sum rule (Eq.~\eqref{sum_rule}). The sum rule has to be treated with care in both the coherent and the strong scattering limit. In the incoherent limit, and specially near mobility edges, we have the relation
$m_{\alpha} = \d{\hbar}{D_{\alpha}}$ which then gives a very
appealing  result, namely
\begin{equation}\label{new_sj_sigxy}
    \sigma_{xy}^{sj}=
    -\d{e^2}{\Omega}\d{\hbar}{4m^2c^2}\sum_{\alpha}f(\varepsilon_{\alpha})\d{mD_{\alpha}}{\hbar}\sigma_{z}^{\alpha}
\end{equation}
This gives a vanishing side-jump-like Hall effect (Rashba) contribution from
those states in the localized part of the band.\\
\\
Adding Eqs.~\eqref{new_sj_sigxy} and Eq.~\eqref{sigxy3} we have the
strong scattering Hall conductivity limit. As in
the QHE, every conducting state contributes. However, the full
enhancement of the spin-orbit coupling (see
Ref.~\onlinecite{DeAndrada} and Ref.~\onlinecite{Engel} for example)
does not appear in our expression. The most interesting aspect of the side-jump-like (Rashba)
term (Eq.~\eqref{new_sj_sigxy}), is that if the Fermi level is in a
region of localized states, say near the top of the band, then this
term dominates the Hall current since all other terms are Fermi
level terms, and vanish at low temperatures. It would give us (using
the free mass) $J_y \sim 10^{-29}N_{del}\langle\sigma_z\rangle F_x$
A/m$^2$ ($N_{del}$ = delocalized electrons density) which is, of
course, a small but \emph{rigid} current, analogous to the Quantum
Hall current.
\section{Conclusion}
The principal result of the cumulant method developed
in this paper is given by Eq.~\eqref{tot_conduc}. It is an expression for
the Hall current which allows for disorder and spin-orbit coupling.
For a typical doped semiconductor, we might expect the following contributions to the Hall effect:
\begin{enumerate}
\item
The normal process (first term of Eq.~\eqref{sigxy3}).
\item
The intrinsic AHE side jump like process (Rashba term,
Eq.\eqref{new_sj_sigxy}).
\item
The intrinsic AHE Karplus-Luttinger process (second term of
Eq.~\eqref{sigxy3}).
\item
The skew scattering process due to impurities
(Eq.~\eqref{skew_sigma}). Here, the basic spin-orbit coupling has
the effective mass and not the free mass.
\end{enumerate}
The skew term (Eq.~\eqref{skew_sigma}) is only relevant if both the
enhancement and the basic spin-orbit coupling
are large and the concentration of impurities high enough (see
Ref.~\onlinecite{Engel} for estimates). The side-jump term is only
important if the effective mass is small or the Fermi level is in a
region of localized states. In most situations with weak disorder,
we expect the Karplus-Luttinger intrinsic term to dominate the
AHE. In the limit of strong scattering, when the use of Bloch functions and uniform lifetimes
is no longer valid, we may replace the effective mass in Eq.~\eqref{tot_conduc} using the concept of quantum diffusivity i.e. put $m_{\alpha}^{-1}\rightarrow \d{D_{\alpha}}{\hbar}$, and use derivatives, as explained in the text.\\
\\
For magneto-optical, Faraday angle measurements, for example, we
need the frequency dependent Hall conductivity. One can show that in
weak scattering, where we use the effective mass for
Eq.~\eqref{sum_rule}, we add a factor $\d{1}{1 + (\omega\tau)^2}$ to
the four main Hall conductivity results listed above. In strong
scattering, where we use the diffusivity to describe
Eq.~\eqref{sum_rule}, $D(\varepsilon)$ is replaced by
$D(\varepsilon,\omega)$.
\begin{acknowledgments}
The authors thank Pr. Arthur Yelon for reading and commenting on the manuscript. L.-F. A. acknowledge
financial support from Pr. A.-M.S Tremblay during the writing of the paper.
\end{acknowledgments}
\appendix
\section{Derivative of the Hamiltonian with respect to the magnetic
field}\label{appen_deriv_of_H}
With $\textbf{A}_i = (0,B_zx_i)$ and $p_y^0 = mv_y^0$, the kinetic part
of the Hamiltonian is
\begin{equation}
T = \sum_i\left[ \d{m(v_i^0)^2}{2} - eB_zx_iv_y^{0,i} +
\d{e^2B_z^2}{2m}x_i^2 \right].
\end{equation}
Thus, with $H = T + V + H_{so}- \d{g\mu_B}{2}B_z\sum_i\sigma_z^{i}$,
we can write, remembering that without the spin-orbit term in
the velocity operator $v_y^i = v_y^{0,i} - \d{eB_z}{m}x_i$,
\begin{equation}\label{H_deriv_1}
\d{\pd H}{\pd B_z} = \sum_i\Bigg[ -ex_iv_y^{i} -
\d{g\mu_B}{2}\sigma_z^{i}\Bigg].
\end{equation}
Thus,
\begin{equation}\label{deriv_H_appendix}
\left\langle \alpha \left|\d{\pd H}{\pd
B_z}\right|\alpha\right\rangle = -e\langle \alpha |xv_y|\alpha\rangle - \d{g\mu_B}{2}\left\langle
\alpha\left|\sum_i\sigma_z^{i}\right|\alpha\right\rangle.
\end{equation}
We now want to transform the LHS of Eq.~\eqref{deriv_H_appendix} to
obtain Eq.~\eqref{deriv_H}. To do so, we just apply the well known
Hellman-Feynman theorem and we obtain $\left\langle \alpha \left|\d{\pd H}{\pd
B_z}\right|\alpha\right\rangle = \d{\pd\varepsilon_{\alpha}}{\pd B_z}$ and thus Eq.~\eqref{deriv_H}. Remember that $v_y$ in the equation does not contains the spin-orbit term of
Eq.~\eqref{vy_so}.
\section{The second order terms in the cumulant expansion}\label{second_order_appen}
Consider now the second order contribution to the Hall conductivity.
This can be written
\begin{equation}\label{sig_xx_second}
    \sigma_{xy}^{\{2\}} =
    \d{e^2}{2!\Omega}\sum_{\alpha,\beta}\langle\alpha
    |x|\beta\rangle\langle\beta |Hv_y-v_yH|\alpha\rangle\d{\pd^2f(\varepsilon_{\alpha})}{\pd\varepsilon_{\alpha}^2}
\end{equation}
where, as previously, $v_y = \d{-i\hbar}{m}\d{\pd}{\pd y} - \d{eB_zx}{m}$,
dropping the spin-orbit current operator. After evaluating the
commutator we are left with three terms
\begin{equation}\label{first_comm}
    i\hbar\d{eB_z}{m}v_x^0,\; \d{i\hbar}{m}\d{\pd V}{\pd y},\; -\d{i\hbar}{m}p_x\sum_n\lambda_n\sigma_z.
\end{equation}
Only the first and third term give significant contributions in this
order. The first term is
\begin{equation}\label{sigma_xy_first_comm}
\begin{split}
    \sigma_{xy,1}^{\{2\}} &= \d{e^2}{2!\Omega}\sum_{\alpha,\beta}\langle\alpha
    |x|\beta\rangle\langle\beta
    |i\hbar\d{eB_z}{m}v_x^0|\alpha\rangle\d{\pd^2f(\varepsilon_{\alpha})}{\pd\varepsilon_{\alpha}^2}\\
    &=
    \d{e^2}{2!\Omega}\d{\hbar^2eB_z}{m}\sum_{\alpha}\left(-\d{1}{2m_{\alpha}}\right)\d{\pd^2f(\varepsilon_{\alpha})}{\pd\varepsilon_{\alpha}^2},
\end{split}
\end{equation}
where we have used the sum rule of Eq.~\eqref{sum_rule} restricted
to the active energy band. For Bloch states restricted to the 8-Kane
bands with spin-orbit coupling, the RHS of Eq.~\eqref{sum_rule}
reproduces the 8-bands $\textbf{k}\cdot\textbf{p}$ calculated
effective masses. To conclude, we note that the cumulant expansion
is a powerful method when the bandwidth is very narrow and $< k_BT$.
At low temperatures, when we use $m_{\alpha} = m^*$, we obtain for Eq.~\eqref{sigma_xy_first_comm}
\begin{equation}\label{sigma_xy_first_comm_low_T}
    \sigma_{xy,1}^{\{2\}} = -\d{e^2}{2!}\d{\hbar}{m}\left( \d{\hbar eB_z}{2m^*}
    \right)\d{\pd\rho(\varepsilon)}{\pd\varepsilon}\Big|_{\varepsilon=\varepsilon_f}.
\end{equation}
Note that the cumulant expansion is not purely an expansion in powers of $B_z$.
There is a term linear in $B_z$ in almost every order. Using the effective mass method,
the linear term in $B_z$ becomes an expansion in powers of the $V_{disorder}/\varepsilon_f$. The convergence is problematic near band edges where we have localization. We neglect the term given by the second term of Eq.~\eqref{first_comm}. The spin-orbit term, given by he third term of Eq.~\eqref{first_comm}, becomes
\begin{equation}\label{sigma_xy_third_comm}
    \sigma_{xy,3}^{\{2\}} =
    \d{e^2\hbar^2}{2!\Omega}\sum_{\alpha}\d{1}{2m_{\alpha}}\Gamma_{\alpha}
    \d{\pd^2f(\varepsilon_{\alpha})}{\pd\varepsilon_{\alpha}^2}\sigma_z^{\alpha}.
\end{equation}
Again, at low $T$ (neglecting other energy dependence under the
integral), we can use
\begin{equation}
    \sum_{\alpha}\d{\pd^2f(\varepsilon_{\alpha})}{\pd\varepsilon_{\alpha}^2}
    = \d{\pd\rho}{\pd\varepsilon}\Big|_{\varepsilon=\varepsilon_f}.
\end{equation}
If we assume that
$\d{\pd\rho}{\pd\varepsilon}\big|_{\varepsilon=\varepsilon_f}\sim
\d{\rho}{\varepsilon_f}$, again, we have the same structure as
before in Eq.~\eqref{y_current}, this time with $\Delta g =
\d{\hbar\langle \Gamma_{\alpha}\rangle}{\varepsilon_f}$. With the
numerator of order $10^{-4}$ eV, this term corresponds to an
effective $g$-shift of $10^{-4}$ which is therefore smaller than the
first order cumulant of its type. But, in general  we remind the reader
that the cumulant expansion, as is also true for the configurationally  decoupled  Kubo formula,    is not accurate near the band edges for reasons of Anderson localisation. When used in this region it can only apply above the mobility edge.
\section{Third order cumulant}\label{third_order_appen}
In the third order cumulant, if we neglect terms of order $B_z^2$ and
$\lambda^2$, we have only the terms
\begin{equation}\label{sigma_xy_third_order}
\begin{split}
    \sigma_{xy}^{\{3\}} &=
    \d{e^2}{3!\Omega}\sum_{\alpha}\d{1}{m_{\alpha}}\d{\pd^3f(\varepsilon_{\alpha})}{\pd\varepsilon_{\alpha}^3}\\ &\left\langle\alpha\left|x^2\nabla_y^2V(\textbf{r})\left[\d{\hbar eB_z}{m} +
    \sum_n\hbar\lambda_n(\textbf{r})\langle\sigma_z\rangle\right]\right|\alpha\right\rangle\d{\hbar}{m}.
\end{split}
\end{equation}
The terms linear in $B_z$ and spin-orbit coupling always go in
pairs, the spin-orbit term acting like an effective magnetic field.
The higher order terms form an infinite series with products
involving derivatives of the lattice potential. These terms are
small at high temperature and renormalize the first order linear
terms in $B_z$ and $\lambda_n$.  In the effective
mass approach, the lattice potential no longer appears in
the Hamiltonian so $V(\textbf{r})$ appearing in Eq.~\eqref{sigma_xy_third_order} is due to impurities or disorder.
\section{Effective mass and diffusivity in the strong scattering limit}\label{strong_scatt}
Since the excited states can decay and must have a finite lifetime,
we can write the sum of Eq.~\eqref{sum_rule} as
\begin{equation}\label{1_m}
    \d{1}{m(\varepsilon_{\alpha})} =
    2\int_{-W}^Wd\varepsilon_{\beta}\rho(\varepsilon_{\beta})\d{ |\langle\alpha |v_x|\beta\rangle|^2(\varepsilon_{\beta}-\varepsilon_{\alpha}) }{ (\varepsilon_{\beta}-\varepsilon_{\alpha})^2 + \left(\d{\hbar}{\tau_{\beta}}\right)^2
    }.
\end{equation}
The quantum diffusivity is defined as
\begin{equation}
    D_{\alpha} = \hbar\sum_{\beta}|\langle\alpha
    |v_x|\beta\rangle|^2\delta(\varepsilon_{\beta}-\varepsilon_{\alpha}).
\end{equation}

Then it follows that we can write
\begin{equation}\label{1_m_1}
\begin{split}
    &\d{1}{m(\varepsilon_{\alpha})} =\\
    &2\int_{-W}^W d\varepsilon_{\beta}\left\{\rho(\varepsilon_{\beta}) \d{ |\langle\alpha |v_x|\beta\rangle|^2\d{\hbar}{\tau_{\beta}}}{ (\varepsilon_{\beta}-\varepsilon_{\alpha})^2 +
    \left(\d{\hbar}{\tau_{\beta}}\right)^2}\right\}\left\{\d{(\varepsilon_{\beta}-\varepsilon_{\alpha})}{\d{\hbar}{\tau_{\beta}}}\right\}\\
    &= \d{2c_{\alpha}D(\varepsilon_{\alpha})}{\hbar},
\end{split}
\end{equation}
where $c_{\alpha}$ is a constant, defined by Eq.~\eqref{1_m_1}. Its
value and sign depends on the density of states variation. One can
also see this by doing an integration by parts using the product of
the two functions in brackets. The constant carries a sign (as does
the effective mass) which depends on the energy. We can rewrite it
as $c_{\alpha} = \left\langle (\varepsilon_{\beta}-\varepsilon_{\alpha})\d{\tau_{\beta}}{\hbar}\right\rangle_{\beta}$,
where the average is to be taken with the diffusivity weighting
function under the integral as defined by Eq.~\eqref{1_m_1}. This equation is still \emph{exact}. In the random phase approximation,
kinetic energies are of the same order as the energy uncertainty and
$c_{\alpha}$ is $\sim  -1$ or $+1$ depending on whether we are near
the top or the bottom of the band.

\end{document}